\documentclass[aip,amsmath,amssymb,reprint,cha]{revtex4-2}

\usepackage{graphicx}
\usepackage{dcolumn}
\usepackage{bm}
\usepackage[utf8]{inputenc}
\usepackage[T1]{fontenc}
\usepackage{mathptmx}
\usepackage{etoolbox}
\usepackage{amsmath}
\usepackage{amssymb}
\usepackage{braket}
\usepackage{bbm}
\usepackage{tikz}
\usepackage{caption}
\usepackage{subcaption}
\usepackage{hyperref}
\hypersetup{
    colorlinks=true,
    linkcolor=blue,
    filecolor=blue,      
    urlcolor=blue,
    citecolor = blue,
}

\allowdisplaybreaks

\makeatletter
\def\@email#1#2{%
 \endgroup
 \patchcmd{\titleblock@produce}
  {\frontmatter@RRAPformat}
  {\frontmatter@RRAPformat{\produce@RRAP{*#1\href{mailto:#2}{#2}}}\frontmatter@RRAPformat}
  {}{}
}%

\makeatother

\newcommand{\revI}[1]{{#1}}
\newcommand{\revII}[1]{{#1}}
\begin{document}
\preprint{AIP/123-QED}

\title[Periodic Orbits in Fermi--Pasta--Ulam--Tsingou Systems]{Periodic Orbits in Fermi--Pasta--Ulam--Tsingou Systems}

\author{Nachiket Karve}
\email{nachiket@bu.edu}
\author{Nathan Rose}
\author{David Campbell}
\affiliation{Department of Physics, Boston University, Boston, Massachusetts 02215, USA}

\date{\today}

\begin{abstract}
    The FPUT paradox is the phenomenon whereby a one-dimensional chain of oscillators with nonlinear couplings shows \revI{long-lived} nonergodic behavior \revI{prior to thermalization}. The trajectory of the system in phase space, with a long wavelength initial condition, closely follows that of the Toda model over short times, as both systems seem to relax quickly to a non-thermal, metastable state. Over longer times, resonances in the FPUT spectrum drive the system towards equilibrium, away from the Toda trajectory. Similar resonances are observed in $q$-breather spectra, suggesting that $q$-breathers are involved in the route towards thermalization. In this article we first review previous important results related to the metastable state, solitons, and $q$-breathers. We then investigate orbit bifurcations of $q$-breathers and show that they occur due to resonances, where the $q$-breather frequencies become commensurate as $m\Omega_1 = \Omega_k$. The resonances appear as peaks in the breather energy spectrum. Furthermore, they give rise to new "composite periodic orbits", which are nonlinear combinations of multiple $q$-breathers that exist following orbit bifurcations. We find that such resonances are absent in integrable systems, as a consequence of the (extensive number of) conservation laws associated with integrability.
\end{abstract}

\maketitle

\begin{quotation}

In 1953, Enrico Fermi, John Pasta, Stanislaw Ulam, and Mary Tsingou (FPUT) studied a one-dimensional chain of oscillators with nonlinear interactions that showed nonergodic behavior \cite{fermi}. Since their original study, many authors have shown that the FPUT system, when initialized in a long wavelength state, goes through a long-lived phase \revI{of recurrent motion, known as the metastable state,} before eventually reaching equipartition \cite{benettin,kevin,Lorenzoni}. \revI{While the system is in the metastable state, the energy spectrum is exponentially localized around the first mode}. Resonant peaks in the energy spectrum are observed to take the system towards equilibrium, albeit very slowly. Although previous attempts have been made to explain this behavior \cite{lvov,flach_tailRes,benettin}, they have not fully captured the resonance structure of the FPUT system. Here, we report on exact resonances between periodic orbits in FPUT systems. These result in the formation of resonant peaks in their energy spectra. Moreover, we show that such resonances give rise to composite periodic orbits, which exist exclusively in the non-integrable FPUT systems and are absent in the integrable Toda lattice and Korteweg-De Vries (KdV) systems.
\end{quotation}

\section{Introduction}

The Fermi--Pasta--Ulam--Tsingou (FPUT) paradox deals with the extremely slow \revI{thermalization} of a chain of anharmonically coupled oscillators initialized far from equilibrium. If the oscillator chain is prepared in a long wavelength mode and the nonlinearity is weak, or equivalently, the energy is small, then there is a long duration interval in which the system undergoes regular recurrences to the initial state \cite{fermi,tuck}. Furthermore, the energy remains exponentially localized in a small number of modes. This long-lived, non-ergodic behavior is known as the metastable state \cite{benettin, kevin, fucito}.  

Dynamics of the FPUT systems have often been understood in relation to nearby integrable nonlinear models. While the original FPUT system was a discrete chain of coupled nonlinear oscillators,  Zabusky and Kruskal (ZK) \cite{kruskal} studied the continuum limit of the so-called $\alpha$-FPUT model and showed it was equivalent to the Korteweg--De Vries (KdV) partial differential equation and contained the celebrated ``soliton" solutions. Indeed, it was in the ZK work that the term ``solitons" was first coined. \revI{The long-wavelength initial state separates into multiple solitons, which move at different speeds and thus drift apart from one another. The eventual re-overlapping of these solitons is responsible for the recurrence phenomenon and recurrence times can be predicted from the speed of the solitons \cite{kruskal}.} A similar analysis has been performed in the $\beta$-model \cite{sal}, whose continuum limit is the modified Korteweg de Vries (mKdV) equation. Alternatively, the Toda lattice is a discrete integrable model close to the discrete $\alpha$-FPUT model \cite{toda,toda2012theory}. The Toda integrals remain approximately conserved during the short-time evolution in the FPUT system, which delays thermalization \cite{BeChPo13}. As an integrable system, the Toda lattice also exhibits exact discrete solitary wave solutions, called cnoidal waves \cite{shastry}.  

A different approach, which we will focus on primarily in this paper, is to consider trajectories of the metastable state as lying close to a special class of exactly periodic orbits of the discrete FPUT systems, known as $q$-breathers \cite{flach}. Generally, a breather is a periodic orbit localized in space, which can be continued from the linear limit. A $q$-breather is likewise a periodic orbit that is localized in normal mode space. The existence of $q$-breathers at sufficiently small nonlinearity \revII{follows from a nonresonance condition of the linear normal mode frequencies \cite{Ly1907, flach}}, and there is a known numerical technique for computing them at finite nonlinearity \cite{flach}. A benefit of this approach is that it does not require taking any continuum limits or introducing different models. As the metastable state also features exponential energy localization, it has been argued that the FPUT trajectory is a perturbation of the exactly periodic $q$-breather orbit \cite{flach,Danieli_2017}. 

As long as the $q$-breather is stable, a trajectory at a sufficiently small but finite distance away from it is expected to remain nearby for a long time. Thus, we investigate how the stability of $q$-breathers changes with the nonlinearity. We find instability regions in both $\alpha$- and $\beta$-FPUT models, \revI{arising when the frequencies of two $q$-breathers become commensurate}. Such resonances result in the creation of new periodic orbits, or "composite periodic orbits", as we call them here. \revI{A Composite periodic orbit can be thought of as a nonlinear combination of two or more $q$-breathers that arises during an orbit bifurcation}. Such periodic orbits are absent in the linear \revI{oscillator chain} due to the incommensurability of the normal mode frequencies. Despite having similar frequency overlaps, these resonances are also absent in the discrete Toda model, due to the extensive number of conservation laws in the completely integrable nature of the Toda model.

The remainder of this paper is organized as follows. In Section \ref{sec_model} we introduce the discrete systems we have examined, namely the $\alpha$- and $\beta$-FPUT models, as well as the Toda lattice. In Section \ref{sec_solitons} we discuss the role of \revI{solitary waves} \cite{Vainchtein} in the dynamics of the FPUT models as well as their relations to true solitons that exist in the continuum limits of these models. In Section \ref{sec_periodicOrbits}, we \revI{review} the numerical procedures for computing $q$-breathers and their stability and present specific examples of $q$-breathers. These first three sections are primarily reviews of prior results that are essential for a full understanding of our study.

The primary novel results of our work appear in Section \ref{sec_resonances}, where we demonstrate the existence of "composite periodic orbits" in non-integrable models that emerge from resonances between $q$-breathers. This sheds light on the complex and diverse nature of the set of periodic orbits in such systems. In section \ref{sec_conclusions} we summarize our conclusions and lay out possible directions for future work. Finally, in Appendix \ref{app_periodic} we discuss the FPUT system with periodic boundary conditions, and in Appendix \ref{appFloquet} we describe the Floquet theory used for the stability analysis of a periodic orbit. A link for codes used for the numerical computations in this paper can be found in Sec. \ref{sec_data}.

\section{The FPUT Model}
\label{sec_model}

The Hamiltonian of a one-dimensional chain of $N$ identical oscillators with nonlinear interactions between nearest neighbors has the general form:
\begin{equation}
H = \sum_{n = 1}^N \frac{1}{2} p_n^2 + \sum_{n=0}^N V(q_{n+1}-q_n).
\label{eq_hamiltonian}
\end{equation}

\revI{FPUT initially investigated models with cubic and quartic interactions, now referred to as the $\alpha$-FPUT and $\beta$-FPUT models, respectively. Their interaction potentials take the form $V_{\text{$\alpha$-FPUT}}(x) = \frac{1}{2}x^2 + \frac{1}{3}\alpha x^3$ and $V_{\text{$\beta$-FPUT}}(x) = \frac{1}{2}x^2 + \frac{1}{4}\beta x^4$. The constants $\alpha$ and $\beta$ capture the strengths of the cubic and quartic nonlinearities. A model that contains both cubic and quartic terms will be referred to as the $\alpha-\beta$-FPUT model. On the other hand, the integrable Toda model has interactions of the form $V_{\text{Toda}}(x) = V_0\left[e^{x/\sqrt{V_0}} - \frac{x}{\sqrt{V_0}} - 1\right]$. Note that when $V_0$ is large, $V(x) \approx \frac{1}{2}x^2 + \frac{1}{6\sqrt{V_0}}x^3 + \mathcal{O}(V_0^{\text{-}1})$, which is equivalent to the $\alpha$-FPUT model with $V_0 = \frac{1}{(2\alpha)^2}$. Therefore, as shown in \cite{Hofstrand, kevin}, the Toda model approximates the dynamics of the $\alpha$-FPUT system at short times. The linear oscillator chain corresponds to the interaction potential $V_{\text{Linear}}(x) = \frac{1}{2}x^2$.}


\revI{We primarily consider lattices with fixed boundary conditions ($q_0 = q_{N+1} = 0$), except when making comparisons with the KdV and mKdV equations in Sec. \ref{sec_solitons}, where we use periodic boundaries (see appendix \ref{app_periodic} for more details on systems with periodic boundary conditions)}. 

The equations of motion for $\alpha$- and $\beta$-FPUT systems are:
\begin{subequations}
    \begin{equation}
        \ddot{q}_n = q_{n+1} + q_{n-1} - 2q_n + \alpha \left[(q_{n+1}-q_{n})^2 - (q_{n}-q_{n-1})^2\right],
        \label{eq_eom}
    \end{equation}
    \begin{equation}
        \ddot{q}_n = q_{n+1} + q_{n-1} - 2q_n + \beta \left[(q_{n+1}-q_{n})^3 - (q_{n}-q_{n-1})^3\right].    
    \end{equation}
\end{subequations}

The normal mode \revI{coordinates} are introduced by the canonical transformation:
\begin{equation}
\begin{pmatrix}Q_k \\ P_k\end{pmatrix} = \sqrt{\frac{2}{N+1}} \sum_{n=1}^N \sin\left(\frac{\pi nk}{N+1}\right) \begin{pmatrix}q_n \\ p_n\end{pmatrix}.
\end{equation}

The Hamiltonians of the $\alpha$- and $\beta$-FPUT models, written in \revI{terms of the normal mode coordinates}, describes $N$ harmonic oscillators with a frequency spectrum of $\omega_k = 2\sin\left(\frac{k\pi}{2(N+1)}\right)$, interacting via many-body couplings:
\begin{subequations}
    \begin{align}
        H_{\text{$\alpha$-FPUT}} = \sum_{k=1}^N\frac{1}{2}\left(P_k^2 + \omega_k^2 Q_k^2\right) + \frac{\alpha}{3}\sum_{i,j,k = 1}^N A_{i,j,k} Q_iQ_jQ_k,
    \end{align}
     \begin{align}
        H_{\text{$\beta$-FPUT}} = \sum_{k=1}^N\frac{1}{2}\left(P_k^2 + \omega_k^2 Q_k^2\right) + \frac{\beta}{4}\sum_{i,j,k,l = 1}^N B_{i,j,k,l} Q_iQ_jQ_kQ_l,
    \end{align}
\end{subequations}

where,
\begin{subequations}
    \begin{equation}
        A_{i,j,k} = \frac{\omega_i\omega_j\omega_k}{\sqrt{2(N+1)}}\sum_{\pm} \left[\delta_{i\pm j\pm k, 0} - \delta_{i\pm j\pm k, 2(N+1)}\right],
    \end{equation}
    \begin{equation}
        B_{i,j,k,l} = \frac{\omega_i\omega_j\omega_k\omega_l}{2(N+1)}\sum_{\pm} \left[\delta_{i\pm j\pm k \pm l, 0} - \delta_{i\pm j\pm k \pm l, \pm 2(N+1)}\right].
    \end{equation}
\end{subequations}

In the absence of nonlinearity, the energy of each normal mode $E_k = \frac{P_k^2 + \omega_k^2Q_k^2}{2}$ is independently conserved. It is the nonlinearity that allows mixing between different modes. \revI{In averaging over typical states of a thermalized system, one should find the energy distributed equally among the normal modes.} We can therefore use the following definition of spectral entropy to measure how close a state is to being thermalized:
\begin{equation}
    S = -\sum_{k=1}^N \epsilon_k \ln \epsilon_k, \text{ where } \epsilon_k = \frac{E_k}{\sum_{k'=1}^NE_{k'}}.
\end{equation}

When all the system's energy is confined to a single mode, the entropy is $S = 0$, whereas the entropy of a thermal state is $S \approx \ln N$. 

All FPUT systems can be scaled to have a total energy of $E = 1$ by the transformation $P_k \to P_k/\sqrt{E}$ and $Q_k \to Q_k/\sqrt{E}$. The nonlinearities transform as $\alpha \to \alpha \sqrt{E}$ and $\beta \to E\beta$ under this scaling. Therefore, the dynamics of an FPUT system can be completely captured by the parameters $E\alpha^2$ and $E\beta$. 

Numerical integration of the equations of motion was done using a symplectic fourth order Runge--Kutta integrator \cite{rk4} with a timestep of $\Delta t = 10^{-4}$. The accuracy of this integrator was tested by performing time reversal tests, as described in \cite{10.1063/1.5079659}. The KdV and mKdV equations were integrated using a regular fourth-order Runge--Kutta integrator. Various other integrators used for nonlinear equations are discussed in \cite{taha}.

\section{Soliton Dynamics}
\label{sec_solitons}

\begin{figure}
    \centering
    \begin{subfigure}[t]{0.47\linewidth}
        \centering
        \includegraphics[width=\linewidth]{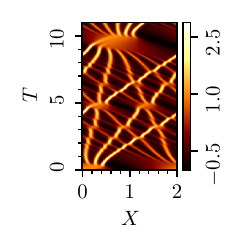}
        \caption{}
    \end{subfigure}
    \hspace{1em}
    \begin{subfigure}[t]{0.47\linewidth}
        \centering
        \includegraphics[width=\linewidth]{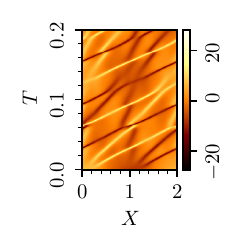}
        \caption{}
    \end{subfigure}
    \caption{Evolution of the KdV (a) and mKdV (b) equations initialized in the state $u(X,0) = A\cos\left(\frac{2\pi x}{L}\right)$. $A = 1$ and $\delta = 0.022$ in (a), while $A = 7.071$ and $\zeta = 0.0278$ in (b). The state breaks apart into multiple solitons which almost overlap with each other again at later times and hence show recurrences of the initial state. \revI{The evolution of the KdV equation is also demonstrated in Video 1 of the supplementary material.}}
    \label{fig_continuum}
\end{figure}

\begin{figure*}
    \centering
    \begin{subfigure}[t]{0.33\linewidth}
        \centering
        \includegraphics[width=\linewidth]{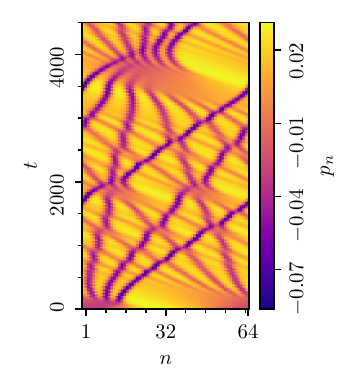}
        \caption{}
    \end{subfigure}
    \hspace{1em}
    \begin{subfigure}[t]{0.63\linewidth}
        \centering
        \includegraphics[width=\linewidth]{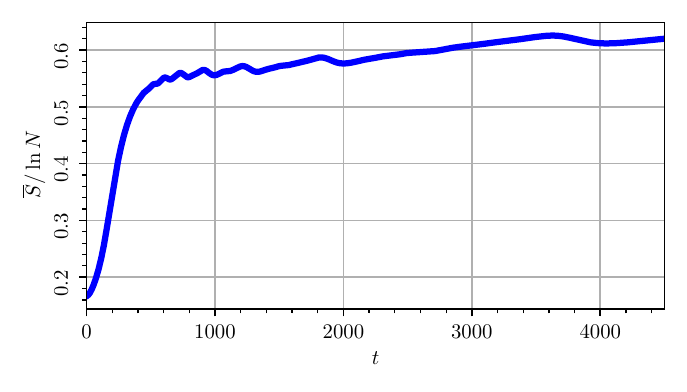}
        \caption{}
    \end{subfigure}
    \begin{subfigure}[t]{0.33\linewidth}
        \centering
        \includegraphics[width=\linewidth]{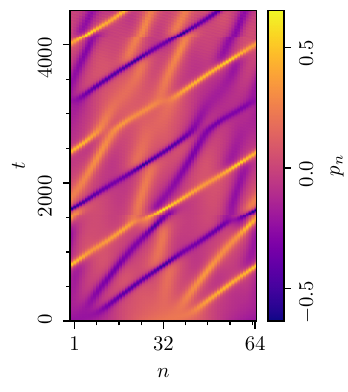}
        \caption{}
    \end{subfigure}
    \hspace{1em}
    \begin{subfigure}[t]{0.63\linewidth}
        \centering
        \includegraphics[width=\linewidth]{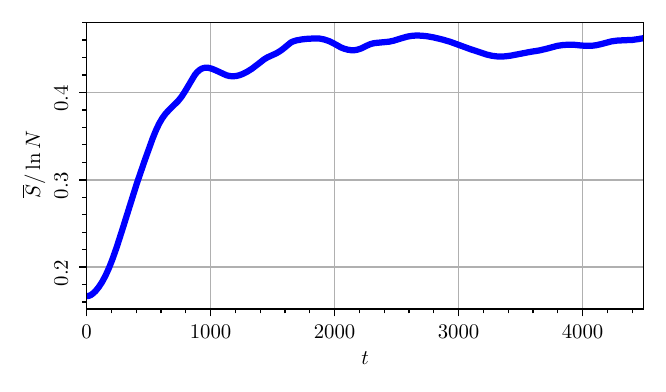}
        \caption{}
    \end{subfigure}
    \caption{(a) Evolution of a 64 particle $\alpha$-FPUT system with energy $E = 0.032$, and nonlinearity $\alpha = 2.69$ viewed in the discrete co-moving frame. The system quickly breaks apart into solitons and forms the metastable state. Note similarities with the KdV system in Fig. \ref{fig_continuum}. (b) Time-averaged spectral entropy of the system $\left(\overline{S}(t) = \frac{1}{t}\int_0^t dt' \ S(t')\right)$, scaled by $\ln N$, as a function of time. The metastable state is achieved when the entropy stabilizes to a non-thermal value. The formation of the metastable state coincides with the breakdown of the system into solitons. (c and d) Evolution of a 64 particle $\beta$-FPUT model with energy $E = 1.55$ and $\beta = 1$.}
    \label{fig_discrete}
\end{figure*}

\begin{figure*}
    \centering
    \begin{subfigure}[t]{0.6\linewidth}
        \centering
        \includegraphics[width=\linewidth]{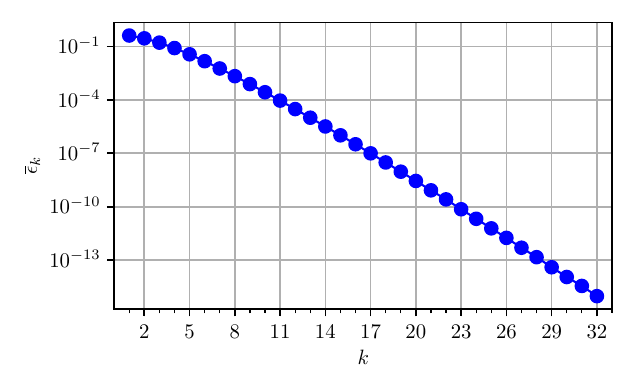}
        \caption{}
    \end{subfigure}
    \hspace{1em}
    \begin{subfigure}[t]{0.33\linewidth}
        \centering
        \includegraphics[width=\linewidth]{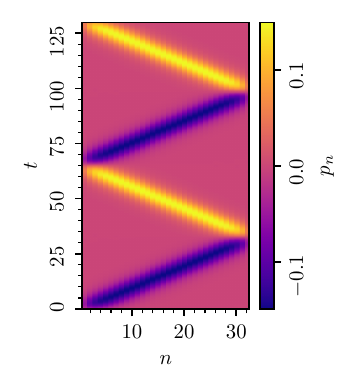}
        \caption{}
        \label{fig_qbreather_mom_a}
    \end{subfigure}
    \begin{subfigure}[t]{0.6\linewidth}
        \centering
        \includegraphics[width=\linewidth]{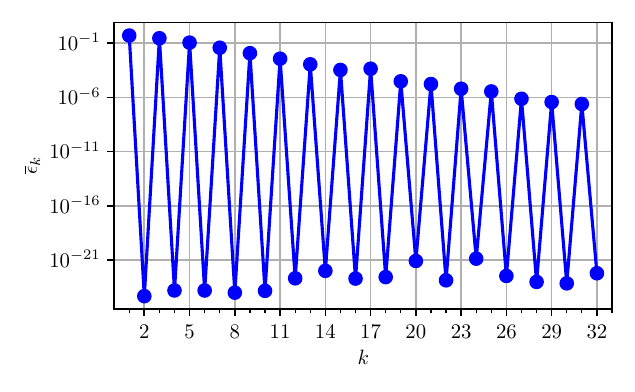}
        \caption{}
    \end{subfigure}
    \begin{subfigure}[t]{0.33\linewidth}
        \centering
        \includegraphics[width=\linewidth]{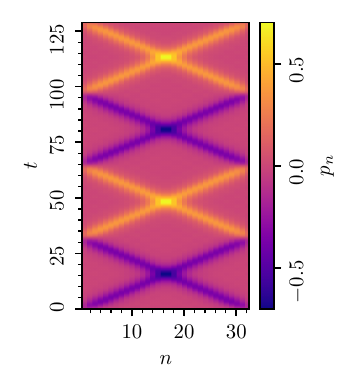}
        \caption{}
        \label{fig_qbreather_mom_b}
    \end{subfigure}
    \caption{$\alpha$-FPUT: (a) Energy as a function of the mode number $k$, averaged over one period of the breather, for a seed mode $k_0 = 1$ $q$-breather in a $32$ particle system with $E\alpha^2 = 4 \times 10^{-3}$. (b) Evolution of its real space momenta. \revI{Video 2 in the supplementary material also demonstrates this evolution.} $\beta$-FPUT: (c) Energy as a function of the mode number $k$, averaged over one period of the breather, for a seed mode $k_0 = 1$ $q$-breather in a $32$ particle system with $E\beta = 0.4$. (d) Evolution of its real space momenta.}
    \label{fig_qbreather}
\end{figure*}

The existence theorem \cite{Friesecke1994} proves the presence of solitary waves in a wide range of FPUT-like lattices. \revI{Strictly speaking, these solitary waves are not solitons due to the non-integrability of the FPUT models. However, the FPUT models are close to integrable models with true soliton solutions. The finite $\alpha$-FPUT chain is a truncation of the integrable Toda lattice, and the continuum limits of the $\alpha$- and $\beta$-models are the KdV and mKdV equations, respectively. We directly compare the evolution of $\alpha$- and $\beta$-FPUT systems and their continuum limits, and show that the dynamics of FPUT systems in the metastable state are largely dominated by solitary waves.} Periodic systems are studied in this section to avoid any complexities introduced by fixed boundaries.  

The equations of motion, Eq. \eqref{eq_eom}, for a periodic $\alpha$-FPUT system ($q_{N+1} = q_0$) composed of particles with mass $m$ and interactions with a spring constant $k$ are:
\begin{align}
    m\ddot{q}_n =& k\left(q_{n+1} + q_{n-1} - 2q_n\right) \notag\\& + \alpha \left[(q_{n+1}-q_{n})^2 - (q_{n}-q_{n-1})^2\right].
\end{align}

We take the limit where the lattice spacing $a$ goes to zero at a constant total length of $(N+1)a = L$. A continuous field $q(x)$ can be defined as one that approximates the lattice: $q(na) = q_n$. Up to fourth order in $a$, the equations of motion can be expressed in terms of spatial and temporal derivatives of $q$ as:
\begin{equation}
    q_{tt} = c^2 q_{xx} + \frac{c^2a^2}{12} q_{xxxx} + c^2\epsilon q_x q_{xx} + \mathcal{O}(a^6),
\end{equation}

where $c = \sqrt{\frac{k}{m}} a$ describes the speed of propagation on the lattice, and $\epsilon = \frac{2\alpha a}{k}$. The nonlinear behavior of the lattice can be better understood by making a transformation to the co-moving frame, $X = x - ct$ and $T = \frac{c\epsilon}{2} t$. Up to first order in $\epsilon$,
\begin{equation}
    q_{XT} + \frac{a^2}{12\epsilon}q_{XXXX} + q_X q_{XX} + \mathcal{O}(\epsilon^2) = 0.
\end{equation}

Defining $u = q_X$, and $\delta^2 = \frac{a^2}{12\epsilon}$, we arrive at the KdV equation \cite{kdv_solitons}:
\begin{equation}
    u_T + u u_X + \delta^2 u_{XXX} = 0.
\end{equation}

Similar manipulations can be made to obtain the mKdV equation in the continuum limit of the $\beta$-FPUT model:
\begin{equation}
    u_T + u^2 u_{XX} + \zeta u_{XXX} = 0,
\end{equation}

where $\zeta = \frac{1}{36\beta}$.

We perform simulations of these continuum equations where, following \cite{kruskal} and \cite{sal}, we start with a long wavelength initial state $u(X,0) = A\cos\left(\frac{2\pi x}{L}\right)$. The KdV and mKdV systems quickly break apart into distinct solitons with varying speeds, as shown in Fig. \ref{fig_continuum}. Recurrence is observed when all of the solitons nearly overlap again. 

Note that in the corresponding FPUT systems, the soliton structure can be seen by studying the evolution of its mechanical momenta since $p(x,t) \approx -mcu(X,T)$. The continuum initial condition translates to $q_n(t=0) = \frac{L}{2\pi}A\sin\left(\frac{2\pi n}{N+1}\right)$, and $p_n(t=0) = -a\sqrt{mk}A\cos\left(\frac{2\pi n}{N+1}\right)$ in the FPUT system, which amounts to initializing the system in the first mode. To view the breakdown of the system into solitons, it is useful to make a transformation to a discrete co-moving frame, defined as:
\begin{equation}
    n \to \left\lfloor n - \frac{(N+1)\omega_1}{2\pi}t \right\rceil \ (\text{mod } N+1), \hspace{1em} t \to t,
\end{equation}

\revI{where $\lfloor x \rceil$ rounds $x$ to the nearest integer.} The breakdown of the FPUT systems into \revI{solitary waves} is almost identical to their continuum counterparts in Fig. \ref{fig_discrete}. After being initialized in the first mode, the system very quickly achieves the metastable state. \revI{The spectral entropy increases rapidly at first, before reaching a plateau and continuing to increase much more slowly as the system reaches the metastable state}. This stage is dominated by highly localized \revI{solitary waves}, which persist for a very long time, resulting in the long lifetime of the metastable state. Unlike the continuum models, \revI{solitary wave} collisions eventually lead the system to equipartition.

\section{Periodic Orbits}
\label{sec_periodicOrbits}

Although states composed of solitary waves in non-integrable FPUT systems are expected to eventually thermalize, there is a set of measure zero in the phase space containing stable, periodic trajectories, comprised of individual solitary waves that preserve their shape indefinitely. Such periodic orbits, or $q$-breathers, can be numerically computed through various iterative procedures \cite{aubry_breathers}. Here, we describe one such procedure based on the Newton method \cite{flach, flach_periodicOrbits, aubry_breathers, flach_book}.

\subsection{Numerical Computation}

All trajectories of the \revI{linear oscillator chain} are either periodic or quasi-periodic and lie on $N$-dimensional tori in the phase space. The Kolmogorov--Arnold--Moser (KAM) theorem \cite{kam} suggests that some of these tori are continuously deformed, but preserved, as a small amount of nonlinearity is added to the system under some non-resonance conditions. Periodic orbits in the linear \revI{oscillator chain} are confined to single normal modes, and as nonlinearity is added, the energy spills over to other nearby modes. Such $q$-breathers remain exponentially localized around a particular mode.

Consider a $q$-breather that is continuously connected to the normal mode $k_0$ in the linear system. We will call $k_0$ the ``seed mode'' of this $q$-breather. The trajectory of this $q$-breather in the phase space will intersect the half-plane $P_{k_0} = 0$, \revI{and} $Q_{k_0} > 0$ exactly once every time period. Thus, the $q$-breather is a fixed point of the corresponding Poincare map, denoted by $\mathcal{F}$. Under the action of $\mathcal{F}$, any point $\mathbf{y}$ on the plane is evolved until its trajectory crosses the half-plane.

Fixed points of $\mathcal{F}$ can be numerically computed using the Newton--Raphson method \cite{suli_newton}. An \revI{initial} guess $\mathbf{y}^{(0)}$ of the $q$-breather solution is first made, which is then iteratively updated by the action of the Newton map $\mathcal{N}$, defined as:
\begin{equation}
    \mathcal{N}[\mathbf{y}] = \mathbf{y} - \left(\mathbbm{1} - \nabla \mathcal{F}[\mathbf{y}]\right)^{-1}\left(\mathbf{y} - \mathcal{F}[\mathbf{y}]\right), \label{eq_newton}
\end{equation}

where the gradient map $\nabla\mathcal{F}[\mathbf{y}]$ evolves the tangent vectors of $\mathbf{y}$. After every successive iteration $\mathbf{y}^{(i+1)} = \mathcal{N}[\mathbf{y}^{(i)}]$, the \revI{relative error} $\rho = \frac{||\mathbf{y} - \mathcal{F}[\mathbf{y}]||}{||\mathbf{y}||}$ is computed, where $|| \mathbf{y} || = \text{max}\left\{y_k\right\}$. The \revI{error} $\rho$ measures how \revI{far} the current solution is from \revI{being a fixed point, and $\rho$ approaches zero if the algorithm converges to a periodic orbit. In practice, this convergence is limited by numerical precision, and iterations are cut off when a finite error threshold is reached. In this work, we have considered the algorithm to have converged when the error $\rho$ is less than $10^{\text{-} 12}$. For reference, the precision of floating point numbers we are using is of the order $10^{\text{-}16}$.}

The Newton method is always linearly stable; that is, a finite domain of convergence around a fixed point always exists. When the Poincare map admits multiple fixed points, their domains of convergence can have fractal-like boundaries. It is, therefore, important to make a good initial guess close to the desired solution. To obtain a $q$-breather with seed mode $k_0$ in a system with nonlinearity $\alpha$, a good initial guess might be a $q$-breather with the same seed mode $k_0$ but in a system with nonlinearity $\alpha-\Delta\alpha$. Thus, $q$-breathers in systems with arbitrary nonlinearities can be obtained by incrementing the nonlinearity in discrete steps, starting from a single normal mode in the linear system. The step size must be small enough such that every breather is in the domain of convergence of the next one.

Additional conditions need to be imposed on $\mathcal{F}$ so that there is a unique fixed point at a given seed mode, energy, and nonlinearity. If we require the $q$-breather solution to be time-reversal symmetric, then there exists a time when all the mode momenta are simultaneously zero. Further, the mode amplitude $Q_{k_0}$ can be determined given all the other mode amplitudes and the energy of the system. This reduces the number of independent variables \revI{needed to specify the $q$-breather to $N-1$ from the original $2N$.}

The gradient map $\nabla \mathcal{F}$ in Eq. \ref{eq_newton} can be computed over these $N-1$ variables by adding perturbations and evolving the linearized equations of motion:
\begin{align}
\delta\ddot{Q}_k = -\omega_k^2 \delta Q_k &- 2\alpha\sum_{i,j=1}^N A_{i,j,k}Q_i\delta Q_j \notag\\& - 3\beta\sum_{i,j,l=1}^N B_{i,j,k,l}Q_iQ_j\delta Q_l.
\end{align}

By requiring the perturbation to point along a constant energy surface in the phase space, $\delta Q_{k_0}$ can be determined in terms of all other components.

We use the algorithm described above to successfully compute $q$-breathers in both $\alpha$- and $\beta$-FPUT systems (see Fig. \ref{fig_qbreather}). As expected, the energy is exponentially localized around the seed mode of the $q$-breather. In the $\alpha$-model, its shape is described by $\overline{\epsilon}_k \approx k^2 \left(\frac{\alpha^2E(N+1)^3\overline{\epsilon}_1}{\pi^4}\right)^{k-1}\overline{\epsilon}_1$ \cite{ponno_2stage_dynamics}. \revI{In the $\beta$-model, only odd modes are excited in the energy spectrum of the seed mode 1 $q$-breather shown in Fig. \ref{fig_qbreather} due to the potential of the $\beta$-model being an even function.} The slope of the energy spectrum depends on the nonlinearity of the system, and with increasing nonlinearity, the energy tends to spread out more in the mode space \cite{flach}.

\begin{figure}
    \centering
    \begin{subfigure}[t]{0.47\linewidth}
        \centering
        \includegraphics[width=\linewidth]{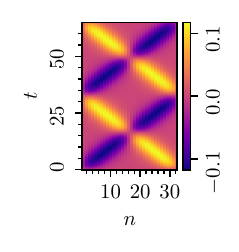}
        \caption{}
    \end{subfigure}
    \hspace{1em}
    \begin{subfigure}[t]{0.47\linewidth}
        \centering
        \includegraphics[width=\linewidth]{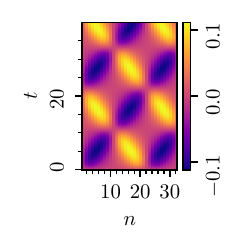}
        \caption{}
    \end{subfigure}
    \caption{Evolution of the momenta of seed mode $k_0 = 2$ (a) and $k_0 = 3$ (b) $q$-breathers in the same system as Fig. \ref{fig_qbreather}. The number of solitons in the breather corresponds to the seed mode number.}
    \label{fig_qbreather_2_3}
\end{figure}

\begin{figure}
    \centering
    \includegraphics[]{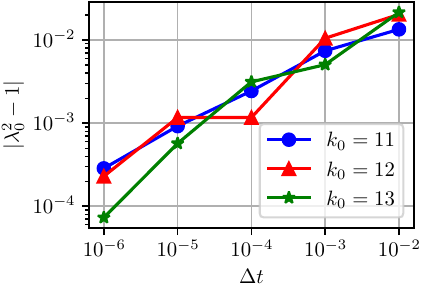}
    \caption{Deviation of the trivial Floquet multiplier from the unit circle as a function of the time step, for seed mode $k_0 = 11, 12, 13$ $q$-breathers in a $16$-particle $\alpha$-FPUT system with $E\alpha^2 = 0.0625$.}
    \label{fig_flqError}
\end{figure}

\begin{figure*}
    \begin{subfigure}[t]{0.475\textwidth}
        \includegraphics[width=\linewidth]{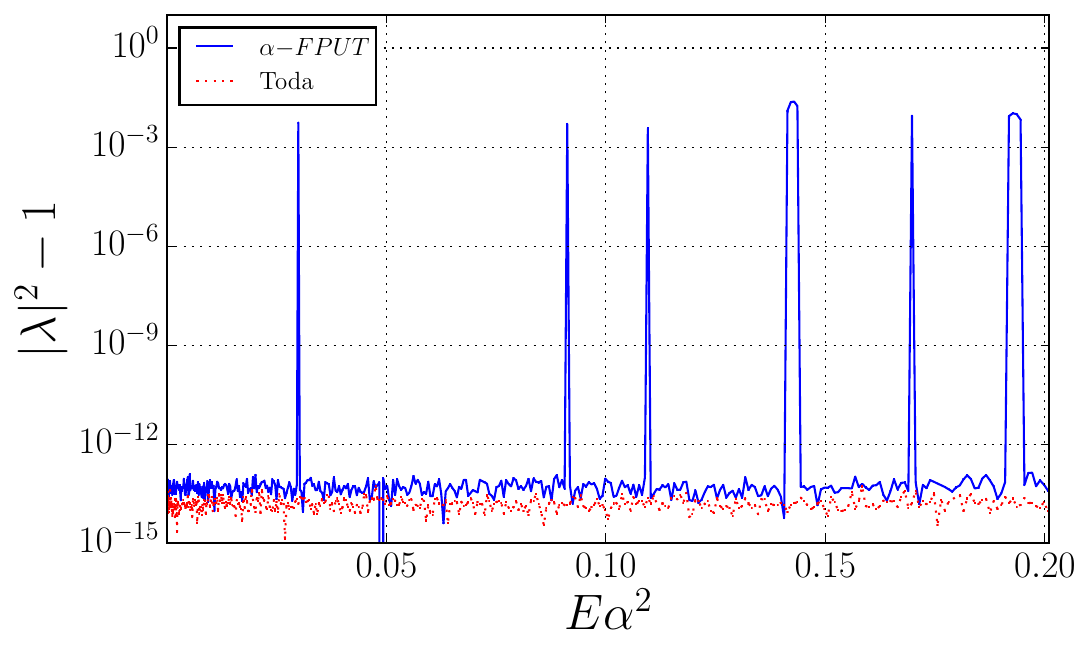}
        \caption{}
    \end{subfigure}
    \begin{subfigure}[t]{0.475\textwidth}
        \includegraphics[width=\linewidth]{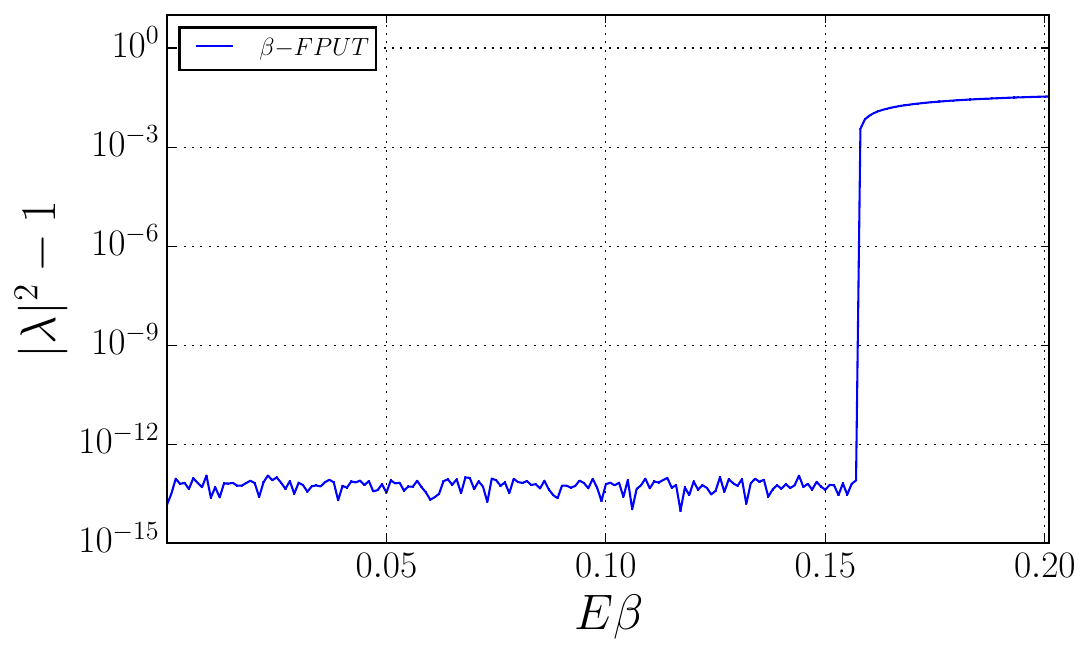}
        \caption{}
    \end{subfigure}
    \caption{Magnitude of the largest non-trivial eigenvalue of the Floquet matrix for computed $q$-breathers in the 16-particle $\alpha\text{-}$FPUT and Toda models (a) and the 16-particle $\beta$-FPUT model (b). Unstable orbits are found in the $\alpha$-FPUT at particular values of nonlinearity, while all orbits found here for the Toda model are stable. Orbits in the $\beta$-FPUT model become and remain unstable above a certain nonlinearity, but remain localized in mode space.
    }
    \label{fig: alpha_toda_stability_comp}
\end{figure*}

\revI{The momentum coordinates of a q-breather trajectory exhibit solitary wave structures localized to a small number of lattice sites, as shown in Figs. \ref{fig_qbreather_mom_a} and \ref{fig_qbreather_mom_b}. These solitary waves pass through one another without deformation, as they must for a periodic orbit. In the $\alpha$-model, a $q$-breather possesses a number of solitary waves equal to its seed mode. In other words, a $q$-breather with seed mode $k_0=1$ contains a single solitary wave, while a $q$-breather with seed mode $k_0>1$ contains $k_0$ equally spaced solitary waves. This is demonstrated in Fig. \ref{fig_qbreather_2_3}, where the momenta as a function of lattice site for $q$-breathers in the $\alpha$-model with seed modes $k_0=2$ and $k_0=3$ are shown.

The solitary waves found in a $q$-breather can be either positive or negative. In the $\alpha$-model, positive waves always traverse the lattice to the left, while negative waves move to the right. In the $\beta$-model positive or negative waves can traverse in either direction, and a $q$-breather contains an equal number of left- and right-moving waves. This situation is analogous to how antisoliton solutions are present in the mKdV equation, but not the KdV equation \cite{sal}.}

\revI{As with solitons, the speed of solitary waves in a $q$-breather are dependent on the energy, or equivalently the strength of nonlinearity. The speed is determined by the $q$-breather frequency, which for small nonlinearity can be found perturbatively as \cite{ponno_2stage_dynamics}}
\begin{equation}
    \Omega_1(E\alpha^2) = \omega_1 + E\alpha^2\frac{\overline{\epsilon}_1\omega_1\omega_2}{8(N+1)(2\omega_1-\omega_2)} + \dots.
\end{equation}
\revI{To summarize, the solitary waves which generally appear in a $q$-breather trajectory seem to be deeply related to solitons in the KdV and mKdV equations.}

\subsection{Stability Analysis}
\label{sec_stb}

The stability of a $q$-breather can be calculated by linearizing the Poincare map around the corresponding fixed point. The $2N$ complex eigenvalues $\{\lambda_i\}$ of the Floquet matrix determine the linear stability of the orbit (see appendix \ref{appFloquet} for more details). \revI{For a Hamiltonian system, the Floquet matrix is symplectic, which implies that its eigenvalues are symmetric with respect to both the unit circle and the real line}. For a $q$-breather to be stable, all of its eigenvalues must lie on the unit circle. We observe that eigenvalues indeed only leave the unit circle in pairs.

\revI{A periodic orbit has an additional constraint on its eigenvalues, which is that there should always be a trivial eigenvalue at $\lambda=1$. The eigenvector corresponding to this eigenvalue is tangent to the trajectory, or in other words it represents an infinitesimal displacement along the direction of the trajectory. For a Hamiltonian system, the symmetry described above implies that there is not one, but a pair of trivial eigenvalues equal to $1$. }

\revI{While there must in theory be a pair of trivial eigenvalues, in practice these eigenvalues may not be exactly equal to 1 due to numerical instabilities \cite{Fairgrieve_okFloq,lust_floq}. We find that in some cases, for instance for a higher seed mode $q$-breather, that the trivial eigenvalues of an otherwise stable periodic orbit leave the unit circle slightly. In Fig. \ref{fig_flqError}, we show the deviation of the eigenvalue at $1$ from the unit circle as a function of the time step $\Delta t$ used in the numerical integration of $q$-breathers with seed modes $k_0=11, 12$, and $13$ and system size $N=16$. Note that at all values of the time step, the deviation is small compared to $1$, but still many orders of magnitude larger than our numerical precision of around $10^{\text{-}16}$. We argue that this is a numerical artifact and does not represent the orbit being truly unstable since the magnitude of instability at $1$ tends to zero as the time step is decreased. Further, we observe that the remaining $2N-2$ non-trivial eigenvalues remain unchanged with varying the time step. Thus, in order to best classify the periodic orbits studied below as stable or unstable, we exclude the pair of eigenvalues around $1$ from our considerations when determining the stability of an orbit. An orbit is determined to be stable if the non-trivial eigenvalues of the Floquet matrix have unit magnitude.}

\revI{Fig. \ref{fig: alpha_toda_stability_comp} depicts the maximum deviation from the unit circle of the non-trivial Floquet multipliers of a $q$-breather as a function of the nonlinearity in both $\alpha$- and $\beta$-FPUT systems. The deviation is never truly zero due to limited numerical precision, but instabilities due to eigenvalue collisions result in clear peaks which extend many orders of magnitude above the baseline deviation of around $10^{\text{-}14}$. Breathers in both models are found to be stable near the linear limit, but demonstrate instabilities at higher nonlinearities. Contrast that with $q$-breathers in the Toda model, which are found to be always stable. }

\section{Composite Periodic Orbits}
\label{sec_resonances}
\begin{figure*}
    \centering
    \begin{subfigure}[t]{0.45\linewidth}
        \centering
        \includegraphics[width=\linewidth]{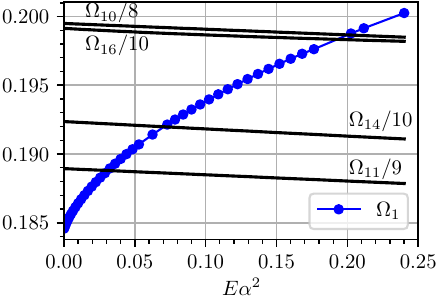}
        \caption{}
        \label{fig_fput_resonances}
    \end{subfigure}
    \begin{subfigure}[t]{0.45\linewidth}
        \centering
        \includegraphics[width=\linewidth]{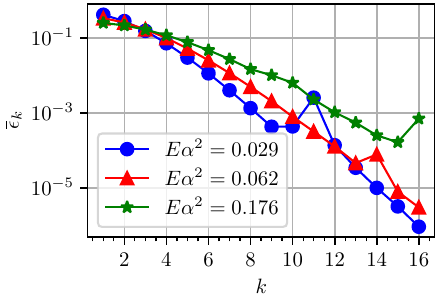}
        \caption{}
        \label{fig_spectra}
    \end{subfigure}
    \begin{subfigure}[t]{0.45\linewidth}
        \centering
        \includegraphics[width=\linewidth]{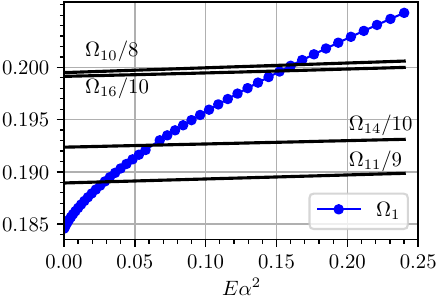}
        \caption{}
        \label{fig_toda_resonances}
    \end{subfigure}
    \begin{subfigure}[t]{0.45\linewidth}
        \centering
        \includegraphics[width=\linewidth]{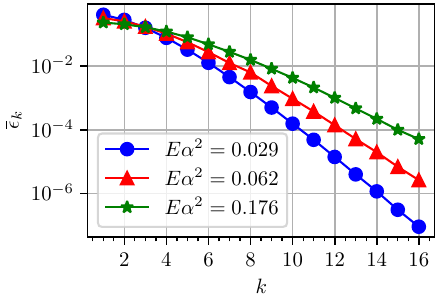}
        \caption{}
        \label{fig_toda_spectra}
    \end{subfigure}
    \caption{(a) Frequency of the seed mode 1 $q$-breather ($\Omega_1$) as a function of the nonlinearity in a 16 particle system, denoted by blue circles. The solid black lines denote fractions of the frequencies of higher mode $q$-breathers. Exact resonances are observed at $E\alpha^2 \approx 0.029$, $0.068$, and $0.194$. (b) The energy spectrum of the seed mode 1 $q$-breather at nonlinearities $E\alpha^2 = 0.029$, $0.062$, and $0.176$, which is close to where the exact resonances occur. Peaks are observed at the resonant modes. (c and d) Corresponding $q$-breathers in the Toda lattice. No resonant peaks are observed despite frequency overlaps.}
    \label{fig_resonance}
\end{figure*}

\revI{The linear oscillator chain has exactly $N$ independent periodic states, corresponding to $N$ normal modes of the system. Any combination of two or more than two linear normal modes necessarily results in a quasi-periodic orbit due to the incommensurability of their frequencies. However, we find that this is not necessarily true for nonlinear non-integrable models such as the $\alpha$- and $\beta$-FPUT models. We show the existence of composite periodic orbits in these models, which emerge from a resonance of two or more that two $q$-breathers. Such composite orbits can be regarded to be a nonlinear combination of multiple $q$-breathers. Interestingly, we do not find such orbits in the Toda model. This leads us to believe that the set of periodic orbits in non-integrable FPUT models is much more complex and diverse than what was previously known.}

\subsection{\texorpdfstring{$q$}{q}-Breather Resonances}

\begin{figure}
    \centering
    \includegraphics[]{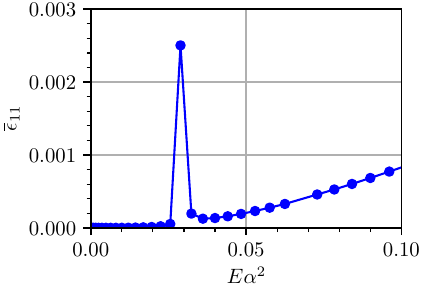}
    \caption{Energy contained in the 11th mode of a seed mode $k_0=1$ $q$-breather in a 16-particle $\alpha$-FPUT system, averaged over one period of the breather, as a function of the nonlinearity. The eleventh mode develops a peak when $9\Omega_1 = \Omega_{11}$, \revI{which is the first resonance observed in Fig. \ref{fig_fput_resonances}}. The peak quickly dies down as we move $E\alpha^2$ away from the resonance.}
    \label{fig_11thMode}
\end{figure}

\begin{figure*}
    \centering
    \begin{subfigure}[t]{0.45\linewidth}
        \includegraphics[width=0.9\linewidth]{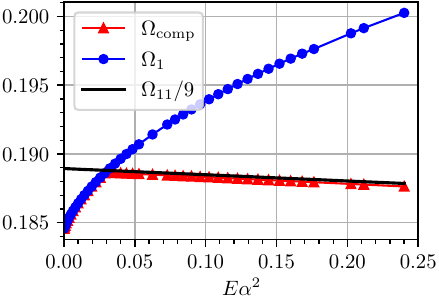}
        \caption{}
        \label{fig_bifurcation_alpha}
    \end{subfigure}
    \begin{subfigure}[t]{0.45\linewidth}
        \includegraphics[width=0.9\linewidth]{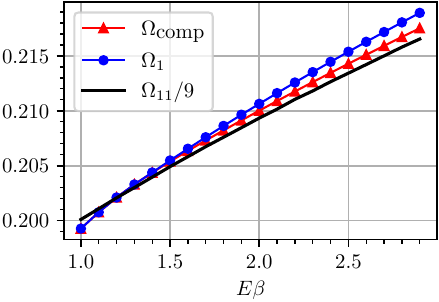}
        \caption{}
    \end{subfigure}
    \caption{Frequencies of composite periodic orbits (red triangles) obtained from a resonance between seed mode $k_0 = 1$ (blue circles) and $k_0 = 11$ (black solid line) $q$-breathers in $\alpha$-FPUT (a) and $\beta$-FPUT (b) systems, both with 16 particles. \revI{The energy spectra of the $q$-breathers and composite orbits in the $\alpha$-model as a function of the nonlinearity are plotted in Video 3 of the supplementary material.}}
    \label{fig_bifurcation}
\end{figure*}

\begin{figure*}
    \centering
    \begin{subfigure}[t]{0.63\linewidth}
        \centering
        \includegraphics[width=\linewidth]{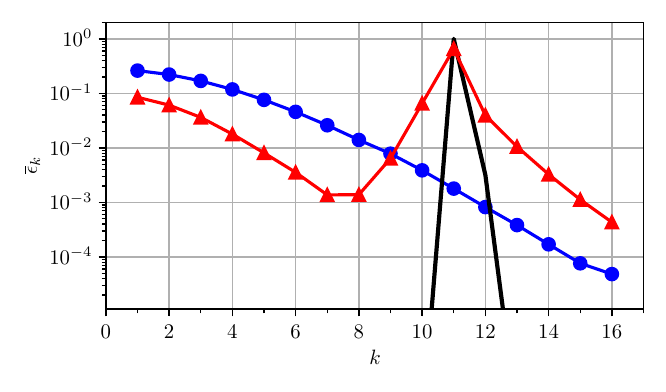}
        \caption{}
        \label{fig_compbr_energy}
    \end{subfigure}
    \begin{subfigure}[t]{0.33\linewidth}
        \centering
        \includegraphics[width=\linewidth]{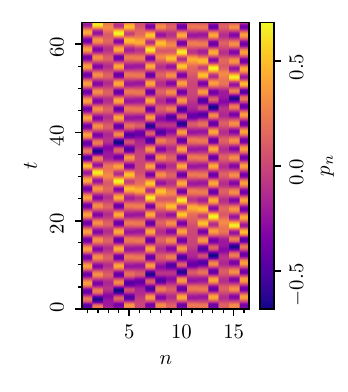}
        \caption{}
        \label{fig_compbr_mom}
    \end{subfigure}
    \caption{(a) The energy spectrum of a composite periodic orbit (red triangles) in a 16 particle $\alpha$-FPUT system with $E\alpha^2 = 0.16$. Modes 1 and 11 contain most of the energy of the breather. The energy spectra of the corresponding seed mode 1 (blue circles) and seed mode 11 (solid black line) $q$-breathers are shown for comparison. (b) Time evolution of the $q$-breather.}
    \label{fig_compbr}
\end{figure*}

\revI{Thermalization of non-integrable systems is often attributed to certain resonance mechanisms. For example, it has been proposed that six-wave resonances in FPUT systems drive a slow diffusion of energy between different modes \cite{lvov}, leading to thermalization of long wavelength initial conditions over long times. The time taken by a long wavelength initial condition to thermalize is predicted to scale as $(E\alpha^2)^{-4}$ and $(E\beta)^{-8}$ in the $\alpha$- and $\beta$-FPUT systems respectively \cite{onorato_waveTurb}. However, numerical experiments suggest that this power law fails in the weakly nonlinear regime, and energy redistribution is instead observed to occur due to resonances between the first mode and higher modes \cite{zhang}. Such resonances are usually identified by the appearance of peaks in the tails of the energy spectrum of the system, which appear to slowly lift up the spectrum towards equipartition \cite{kevin,benettin}. Similar peaks are observed in $q$-breather spectra, suggesting that $q$-breather resonances play some role in the thermalization of FPUT systems \cite{flach_tailRes}.}


Here, we show that peaks in FPUT $q$-breather spectra are manifestations of exact resonances between seed mode $1$ and higher seed mode $q$-breathers. These resonances are of the type $m\Omega_1 = \Omega_k$, where $\Omega_k$ is the frequency of a higher seed mode $q$-breather at the same energy and nonlinearity, and $m$ is some integer. We observe such peaks only in non-integrable systems. Fig. \ref{fig_resonance} compares $\alpha$-FPUT and Toda $q$-breathers at various nonlinearities. Seed mode 1 breathers are numerically computed using the Newton algorithm in a 16-particle $\alpha$-FPUT system, and their frequencies ($\Omega_1$) are plotted as a function of $E\alpha^2$ in Fig. \ref{fig_fput_resonances}. We observe that the frequency monotonically increases with the nonlinearity. In the same figure, we plot fractions of the frequencies of certain higher mode $q$-breathers, $\Omega_k/m$. \revI{The integer $m$ is chosen such that there is an intersection between the curves of $\Omega_1$ and $\Omega_k/m$ in the range of nonlinearity that is being studied. We note that this results in a unique choice of $m$ for each curve in Figs. \ref{fig_fput_resonances} and \ref{fig_toda_resonances}. Interestingly, no suitable choice of $m$ exists for seed modes $k < 8$ in the range of nonlinearity under study.} A resonance between mode 1 and mode $k$ is observed when the two curves intersect. $q$-Breather spectra computed at nonlinearities near such frequency overlaps have peaks at the relevant resonant modes (Fig. \ref{fig_spectra}). The peak dies down quickly as we move away from the resonance (Fig. \ref{fig_11thMode}). On the other hand, $q$-breather spectra in the Toda model contain no resonant peaks despite having similar frequency overlaps (Figs. \ref{fig_toda_resonances} and \ref{fig_toda_spectra}). This suggests that $q$-breather resonances are \revI{suppressed} in integrable systems.

\subsection{Bifurcations}

\begin{figure}
    \centering
    \includegraphics[]{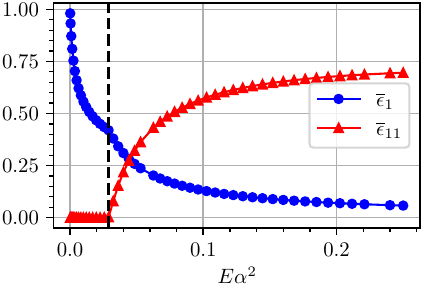}
    \caption{Energies contained by modes 1 and 11 of a composite periodic orbit in a 16-particle $\alpha$-FPUT system, averaged over one period of the orbit, as a function of the nonlinearity. The composite orbit emerges from a resonances between the 1st and 11th modes at $E\alpha^2 \approx 0.029$, denoted by a black dashed line. Note that before the bifurcation the periodic orbit is a seed mode 1 $q$-breather and most of its energy is contained in the 1st mode. However, it becomes a composite orbit after the bifurcation and a rapid exchange of energy occurs between the 1st and 11th modes.}
    \label{fig_energyShare}
\end{figure}

\begin{figure}
    \centering
    \includegraphics[]{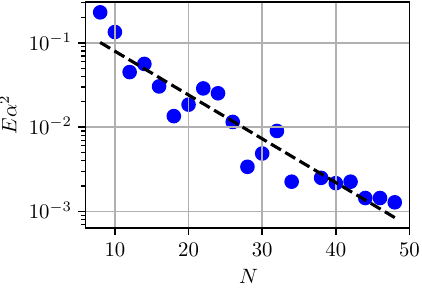}
    \caption{Nonlinearity at which a composite periodic orbit is first encountered as a function of the system size in the $\alpha$-FPUT model.}
    \label{fig_NDependence}
\end{figure}

\begin{figure}
\begin{subfigure}[t]{0.47\linewidth}
    \centering
    \includegraphics[]{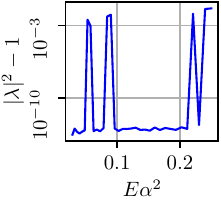}
    \caption{}
\end{subfigure}
\begin{subfigure}[t]{0.47\linewidth}
    \centering
    \includegraphics[]{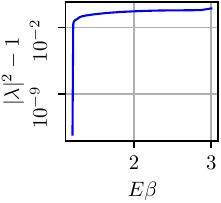}
    \caption{}
\end{subfigure}
\caption{Maximum deviation from the unit circle of the non-trivial Floquet multipliers as a function of the nonlinearity for composite periodic orbits in 16-particle $\alpha$-FPUT (a) and $\beta$-FPUT (b) systems, resulting from a resonance between seed mode 1 and 11 $q$-breathers.}
\label{fig_compStb}
\end{figure}

\revI{Resonances between $q$-breathers of an FPUT system can lead to the formation of a new composite periodic orbit. This composite orbit is a nonlinear combination of the resonating breathers. An orbit that emerges from a resonance $m_1\Omega_1 + \dots m_N\Omega_N = 0$ will be referred to as a composite orbit of the type $(m_1,\dots, m_N)$. Interestingly, it is possible to continue this composite orbit to higher nonlinearities, even when the original $q$-breathers stop resonating. This results in a bifurcation and the composite orbit continues to exist on a separate branch, simultaneously with the original breathers. See Fig. \ref{fig_bifurcation}, where a composite branch arises in both $\alpha$- and $\beta$-FPUT systems due to a resonance between the first and the eleventh mode. Again, we do not observe composite branches in the Toda model.

Numerical computation of the composite orbit can be a bit tricky. At the point of bifurcation the Poincare map $\mathcal{F}$ admits two fixed points: one corresponds to the seed mode 1 breather, while the other corresponds to the composite orbit. To obtain both solutions we perform two separate runs of the Newton algorithm with two different step sizes in the nonlinearity. In general, we observe that the numerical procedure converges to the composite orbit when a smaller step size is used, while a larger step size results in the seed mode 1 breather (although both runs converge to the same breather solution before the bifurcation). Typically, for the $\alpha$-model, we have used values of $\Delta \alpha = 10^{-4}$ to obtain a composite orbit, and $\Delta \alpha = 10^{-2}$ to obtain a breather. The composite orbit can then be continued to larger nonlinearities by using the same smaller step size in the Newton algorithm (see Fig. \ref{fig_bifurcation}).

An example of a composite orbit in a 16-particle $\alpha$-FPUT system, with $E\alpha^2 = 0.16$, is shown in Fig. \ref{fig_compbr}. This orbit is obtained by continuing along the bifurcation that arises due to a resonance between seed mode 1 and 11 breathers in the system at $E\alpha^2 \approx 0.029$ (Fig. \ref{fig_bifurcation_alpha}). The composite orbit retains some features of the original $q$-breathers, comprising of one prominent and eleven smaller solitary waves (Fig. \ref{fig_compbr_mom}). Unlike $q$-breathers, the energy of a composite orbit is not localized around a single mode, but instead, most of the energy is shared between modes 1 and 11 (Fig. \ref{fig_compbr_energy}). In fact, with increasing nonlinearity, mode eleven's energy share in the composite orbit also increases (Fig. \ref{fig_energyShare}).}

\revI{Composite periodic orbits become more abundant with increasing system size. To see this, we study bifurcations in various $\alpha$-FPUT models. As mentioned above, two runs of the Newton algorithm with two different step sizes are performed. A bifurcation is identified when the runs converge to two distinct solutions. Fig. \ref{fig_NDependence} plots the smallest nonlinearity at which a bifurcation is observed as a function of the system size in the $\alpha$-FPUT model, and we observe that this nonlinearity decreases exponentially with increasing $N$. This is expected, since a larger system has a larger number of $q$-breathers, leading to more ways in which breathers can be combined. We speculate that composite orbits are found at all nonlinearities in the $N \to \infty$ limit.}

\subsection{Stability Analysis}

\revII{Following Sec. \ref{sec_stb}, the stability of composite periodic orbits is studied by measuring the deviation of their non-trivial Floquet multipliers from the unit circle. As shown in Fig. \ref{fig_compStb}, composite orbits in the $\alpha$-model are mostly stable, except for certain regions of instability. On the other hand, composite orbits in the $\beta$-model, just like their $q$-breather counterparts, show large instabilities.}

\section{Conclusions}
\label{sec_conclusions}

In this article, we have examined the resonant structure of $\alpha$- and $\beta$-FPUT models. Starting with a comparison between the FPUT metastable state and the soliton structure of the KdV and mKdV equations, we highlight the importance of studying solitary waves to understand thermalization in FPUT systems. Next, a method for obtaining periodic orbits, or $q$-breathers, in FPUT systems is reviewed. \revI{We then argue that there is a direct} connection between $q$-breathers and solitary waves.

Seed mode $k_0=1$ breathers are observed to develop instabilities at certain nonlinearities, corresponding to exact resonances between seed mode $k_0=1$ and higher seed mode q-breathers, $m\Omega_1 = \Omega_k$. These manifest as resonant peaks in the $q$-breather spectrum. Previous works have noted the similarities between such resonant peaks and those observed in the metastable state \cite{flach_tailRes}. Resonances also lead to bifurcations of $q$-breather orbits, resulting in new composite periodic orbits with multiple prominent modes. Composite orbits can be regarded as nonlinear combinations of different breathers. Such orbits are shown to exist at higher nonlinearities, even when the original $q$-breathers stop resonating. Interestingly, such a resonant mechanism is absent in the integrable Toda lattice.

Further work needs to be done to understand how breather resonances affect the dynamics of solitary waves in the system. It will be interesting to study soliton scatterings at nonlinearities close to where resonances occur. Interactions between solitary waves of $q$-breathers and composite orbits \revI{have not previously been systematically studied}. Furthermore, we speculate the existence of other forms of resonances, such as $m_k\omega_k = m_{k'}\omega_{k'}$. These could also result in new periodic orbits via period multiplying bifurcations. This would \revI{allow us to numerically compute a larger class of periodic orbits related to $q$-breather orbit bifurcations.} How these orbits affect the system's path to equilibrium is an open question.

\section{Supplementary Material}

This section describes the videos included in the supplementary material. Video 1 shows the evolution of the KdV equation from Fig. \ref{fig_continuum}. A timestep of $\Delta t = 10^{-4}$ is used. Video 2 shows the evolution of a seed mode $k_0=1$ $q$-breather in a 32-particle $\alpha$-FPUT system with $E = 0.1$ and $\alpha = 0.2$. A timestep of $\Delta t = 10^{-3}$ is used for the simulation. In video 3, the energy spectra of seed mode $k_0 = 1$ $q$-breathers and the energy spectra of composite periodic orbits arising from a resonance between modes $1$ and $11$ in a 16-particle $\alpha$-FPUT system are plotted with changing $E\alpha^2$. The resonance gives rise to a peak at mode $11$ and a bifurcation is seen to occur at $E\alpha^2 \approx 0.029$.

\section{Data Availability}
\label{sec_data}
The authors can provide all data related to the results upon request. Simulation codes used in the paper can be found here: \url{https://github.com/nachiketkarve/fputFixed}. 

\appendix

\section{Periodic Systems}
\label{app_periodic}

Here we describe a system with the Hamiltonian (Eq. \ref{eq_hamiltonian}), but with the boundary condition $q_{0} = q_{N+1}$. The normal mode space is now complex, and is defined by the canonical transformation:
\begin{equation}
    Q_k = \frac{1}{\sqrt{N+1}}\sum_{n=0}^N q_n e^{-i\frac{2\pi kn}{N+1}}.
\end{equation}

The condition that the real space variables must be real implies that $Q_k^* = Q_{N+1-k}$, and $P_k^* = P_{N+1-k}$. The normal mode space Hamiltonian for the $\alpha-\beta$-FPUT system takes the form:
\begin{align}
    H = \sum_{k=1}^N \frac{|P_k|^2 + \omega_k^2 |Q_k|^2}{2} &+ \frac{\alpha}{3}\sum_{i,j,k = 1}^N A_{i,j,k} Q_iQ_jQ_k \notag\\&+ \frac{\beta}{4}\sum_{i,j,k,l = 1}^N B_{i,j,k,l} Q_iQ_jQ_kQ_l,
\end{align}

with frequencies $\omega_k = 2\sin\left(\frac{\pi k}{N+1}\right)$, and couplings:
\begin{subequations}
    \begin{equation}
        A_{i,j,k} = i \frac{\omega_i\omega_j\omega_k}{\sqrt{N+1}} \left[\delta_{i+j+k, N+1} - \delta_{i+j+k, 2(N+1)}\right],
    \end{equation}
    \begin{align}
        B_{i,j,k,l} = -\frac{\omega_i\omega_j\omega_k\omega_l}{N+1} \big[\delta_{i+j+k+l,N+1} &- \delta_{i+j+k+l,2(N+1)} \notag\\& + \delta_{i+j+k+l,3(N+1)}\big].
    \end{align}
\end{subequations}

The linear mode energies are given by $E_k = \frac{|P_k|^2 + \omega_k^2 |Q_k|^2}{2}$. Note that $E_{N+1-k} = E_k$.

\section{Floquet Theory}
\label{appFloquet}

Since $q$-breathers are periodic in time, one can use Floquet theory \cite{Brown2013} to study their stability. Floquet theory deals with linear differential equations of the form:
\begin{equation}
\dot{\mathbf{x}}(t) = \mathbf{A}(t)\mathbf{x}(t),
\label{eqFloquet}
\end{equation}

where $\mathbf{A}(t)$ is an $N\times N$ periodic matrix with a period $T$. This equation can have $n$ linearly independent solutions of the form:
\begin{equation}
\mathbf{x}_i(t) = e^{\mu_i t}\mathbf{p}_i(t),
\end{equation}

where $\mathbf{p}_i(t)$ is periodic with a period of $T$. $\lambda_i = e^{\mu_i T}$s are called the characteristic multipliers of the system of equations and $\mu_i$s are called the Floquet exponents. Further,
\begin{equation}
\mu_1 + \mu_2 + \dots + \mu_n = \frac{1}{T}\int_0^T dt' \ \text{tr}\left(\mathbf{A}(t')\right).
\end{equation}

Note that any solution $\mathbf{x}_i(t)$ gains a phase every period:
\begin{equation}
\mathbf{x}_i(t + nT) = e^{n\mu_iT}\mathbf{x}_i(t).
\end{equation}

Thus, the stability of the solution is determined by the Floquet multiplier:
\begin{itemize}
\item $\left|e^{\mu_iT}\right| < 1$: The solution goes to zero as $t\to\infty$,
\item $\left|e^{\mu_iT}\right| = 1$: The solution is stable,
\item $\left|e^{\mu_iT}\right| > 1$: The solution blows up as $t\to\infty$.
\end{itemize}

To analyze the stability of $q$-breathers in our system we introduce infinitesimal perturbations to the normal modes: $Q_k \to Q_k + \delta Q_k$, $P_k \to P_k + \delta P_k$. These perturbations obey the following linearized equations of motion:
\begin{subequations}
\begin{equation}
\delta\dot{Q}_k(t) = \delta P_k(t),
\end{equation}
\begin{align}
\delta\dot{P}_k(t) = -\omega_k^2 \delta Q_k(t) &- 2\alpha\sum_{i,j=1}^N A_{i,j,k}Q_i(t)\delta Q_j(t) \notag\\& - 3\beta\sum_{i,j,l=1}^N B_{i,j,k,l}Q_i(t)Q_j(t)\delta Q_l(t).
\end{align}
\end{subequations}

\begin{figure}
    \centering
    \begin{subfigure}[t]{0.47\linewidth}
        \centering
        \includegraphics[width=\linewidth]{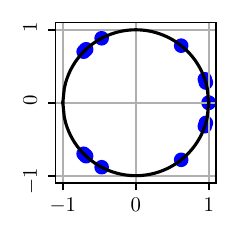}
        \caption{}
    \end{subfigure}
    \begin{subfigure}[t]{0.47\linewidth}
        \centering
        \includegraphics[width=\linewidth]{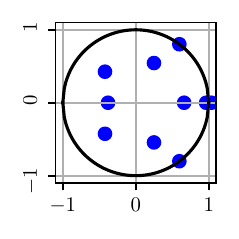}
        \caption{}
    \end{subfigure}
    \caption{Floquet multipliers of a stable seed mode $k_0=1$ $q$-breather (a) and an unstable composite periodic orbit (b) in an 8-particle $\alpha$-FPUT system.}
    \label{fig_floq}
\end{figure}

The above equations are linear and periodic, as required by Eq. \ref{eqFloquet}. The Floquet multipliers of the system can be computed by evolving the perturbations over one period of the breather. Since we are dealing with Hamiltonian systems, there are additional symmetries that are imposed on the Floquet multipliers: for every Floquet multiplier $\lambda_i$, both $\lambda_i^*$ and $1/\lambda_i$ are also Floquet multipliers. Therefore, a $q$-breather is stable if and only if all the Floquet multipliers lie on the unit circle. Examples of stable and unstable breather Floquet multipliers are given in Fig. \ref{fig_floq}.

\bibliography{references}

\end{document}